\documentclass[twocolumn,showpacs,preprintnumbers,amsmath,amssymb,prl]{revtex4}
\usepackage{graphicx}

\begin{document}

\title[]{Epitaxial growth under oblique incidence}%
\author{Z. Moktadir}%
\address{School of Electronics and Computer Science,
Southampton University, Southampton SO17 1BJ, United Kingdom
}%
\email{zm@ecs.soton.ac.uk}%

\thanks{}%

%\date{}%
%\dedicatory{}%
%\commby{}%
% ----------------------------------------------------------------
\begin{abstract}
A continuum one dimensional model of homoepitaxial growth under
oblique incidence is investigated. We carried out numerical
integration of a continuum equation incorporating a shadowing
search algorithm. The interplay between the Ehrlich-Schwoebel (ES)
effect and shadowing is clearly highlighted. It was found that
different growth phases are separated by a well defined crossover
time after which the ES and shadowing are the dominating
mechanisms. Also, the model identified the existence of a
transition period where deep grooves develop on the surface before
the shadowing regime is fully developed.  We found that growth
under oblique angles accelerates the coarsening of mounds on the
surface and we determined the corresponding critical exponent.

\end{abstract}
\maketitle
% ----------------------------------------------------------------
%\section{Introduction}
Epitaxial growth is a process where atoms are deposited on a
crystalline substrate to undergo diffusion on terraces or at the
edge of atomic steps as well as attachment and detachment from
step edges. These microscopic processes determine the surface
morphology of the growing film\cite{Krug02,Villain}. It is well
known that the formation of mounds during Homoepitaxial growth is
a result of asymmetric attachment-detachment kinetics i.e. at the
edge of an
 atomic step, atoms have to overcome an energy barrier to move
downward. This energy barrier is called the Ehrlich-Schwoebel (ES)
barrier\cite{Ehrlish,Schwoebel}. In spite of the fact that many
experimental observations of the surface morphology of films grown
by molecular beam epitaxy have been explained by the above
relaxation processes, it was recently reported that geometrical
effects can play an important role \cite{VanDijken,Amar} when
growth is performed under oblique incidence. In Cu/Cu(100) epitaxy
for example, it was reported that large incidence angles $\theta$
of incoming atomic flux result in the formation of asymmetric
mounds for $\theta=70^o$ and asymmetric ripples on the surface for
$\theta=80^o$ \cite{VanDijken}.  In addition, the crystallographic
orientation of the mounds/ripples facets depend strongly on
temperature. These observations were confirmed by computer
simulations which reveal that they are purely determined by
geometrical effects\cite{Amar}.\\
\indent  Many theoretical aspects of growth in which shadowing
plays an important role were introduced through continuum
models\cite{Karunasiri,Bales}. An example, is the sputter
deposition of films where the observed surface morphologies were
attributed to the influence of shadowing which induces an
instability in the growth process\cite{Bales}. A simple continuum
model that describes this effect is proposed in \cite{Karunasiri},
in which the smoothening effect of diffusion is included as well.
In this model, the height $h$ of the surface grows according to
the following equation:
\begin{equation}\label{eq2}
    \frac{\partial h}{\partial t}=-K\nabla^4h+F\theta + \eta(x,t)
\end{equation}
Model (\ref{eq2}) assumes low vapor pressure and ignores local
non-linearity caused by lateral growth. The parameters $F$  and
$K$ represent the deposition rate and the diffusion process
respectively, while $\eta$ is Gaussian white noise satisfying
$<\eta(x,t)\eta(x',t')>=D\delta(x-x')\delta(t-t')$. The angle
$\theta(x,\{h\})$ is the exposure angle which depends on the
position and the entire surface height $h$. In this example, the
resulting surface has a columnar structure with deep grooves
between neighboring columns, with a well defined characteristic
thickness. The tops of the columns are
growing with a constant speed along the surface normal. \\
\indent Despite these efforts, many aspects of epitaxial growth
under oblique incidence remain unclear. Furthermore, the shadowing
effects have been largely unexplored experimentally due to a lack
of theoretical predictions of the experimentally observable
quantities that discriminate different growth regimes.
 Here we propose a one dimensional continuum model of
homoepitaxial growth under oblique incidence that  incorporates
the ES effect, diffusion and shadowing. We clearly uncover the
interplay between these effects. We show a clear distinction
between the linear regime and the non-linear regime where the
shadowing becomes significant, through the identification of a
well defined crossover time between the two regimes. We identify
the existence of a transition period where deep grooves develop on
the surface before the shadowing regime is fully developed.
Finally, we found that growth at oblique incidence accelerates the
coarsening of mounds
on the surface and we thereafter determined the corresponding critical exponent.\\
%\section{The model}
A phenomenological continuum model describing the surface growth
incorporating the ES effect, diffusion and shadowing can be
formulated in one dimension by:
\begin{equation}\label{fullmbeeq}
    \frac{\partial h}{\partial t}=-\nabla { j_d}-\nabla {
    j_s} +S(\theta,x,h,m)F + \eta(x,t)
\end{equation}
 where $h$ is the single valued surface height, $j_d$ is the ES destabilizing current, $j_s$ is
Mullins stabilizing diffusion current, $F$ is the deposition rate
and $m \equiv \partial h/\partial x$ is the slope.
 A model for the currents $j_d$ and $j_s$ can be expressed as\cite{Krug02,Politi2000,Stroscio}:
 \begin{eqnarray}\label{currents}
    j_d(m)&=& \nu f\left(\frac{m}{m_0}\right)m  \nonumber\\
    j_s(m)&=&-K\nabla^2(m)
 \end{eqnarray}
 Here, $\nu$ and $K$ are positive constants, and the current
 function $f$ satisfies $f(x)=1-x^2$.
  Without the shadowing term equation (\ref{fullmbeeq}) predicts a
  mound-like profile where the symmetric mounds are formed during the initial
  stage of the growth as a result of the competition between the ES
  effect and surface diffusion. Later in time, coarsening of mounds
  occurs with slope selection i.e. the slopes of mounds converge to
  a single value $m_0$\cite{Politi98,Krug02}. \\
The function $S$ is a non-local shadowing term which in general
depends on the incidence angle $\theta$ (which is the angle
between the incoming flux and the y-axis) of the incoming flux,
the surface height and the surface slope at a given position on
the surface profile. It can be defined as the probability that a
point $M(x,h,m)$ of the evolving profile receives the incoming
flux which hits the surface under oblique incidence. This function
arises in the field of optics and was extensively studied to take
into account the shadow effect during the scattering of
electromagnetic waves on randomly rough surfaces( see for example
\cite{Olgilvy} and references therein). The general expression for
the {\it shadowing function} $S$ is given by\cite{Wagner_Smith}:
\begin{equation}\label{shadowingfunc}
    S(\theta, M)=H(\cot\theta-m)\exp\left( -\int_0^\infty
    f(\theta|M,x)dx\right )
\end{equation}
where $H$ is the Heaviside function i.e. $H(x)=0$ for $x\leq0$ and
$H(x)=1$ for $x>0$, and M is a point on the profile having the
height $h$ and the slope $m$. The function $f(\theta|M,x)$ is the
conditional probability that the incoming flux intersects the
surface in the interval $[x,x+dx]$ with the knowledge that it does
not hit the surface in the interval $[0,x]$. This function is
expressed as:
\begin{equation}\label{conditionalprob}
   f(\theta|M,x)=\int_\mu^\infty(m'-\mu)p(h';m'|h;m)dm'
\end{equation}
where $\mu=\cot\theta$ and $h'=h+\mu x$. The function $p$ is the
joint probability distribution of the heights $(h,h')$ and the
slopes $(m,m')$. Analytical expressions of $S$ are possible to
derive for static rough surfaces; for example Smith and Wagner
\cite{Wagner_Smith} and recently Bourilier and Berginc
\cite{Bourlier2003} derived $S$ for Gaussian one
dimensional profiles.\\
For evolving profiles, analytical expressions for $S$ are
intractable if not impossible to obtain, especially when
non-linearities are present. In the latter case the joint
probability in (\ref{conditionalprob}) may not be the product of
individual height and slope probabilities. It is however possible
to compute numerically and with great accuracy this function for
any profile, using a numerical algorithm \cite{Brokelman}. This
purely geometric algorithm determines if a point in the profile is
shadowed or illuminated. We use this algorithm to determine $S$ in
(\ref{fullmbeeq}). In our model, values of $S=1$ are attributed to
illuminated areas while values of  $S=0$ are attributed to
shadowed areas.
%\section{Results and discussion}
\begin{figure}
  \includegraphics[width=10cm]{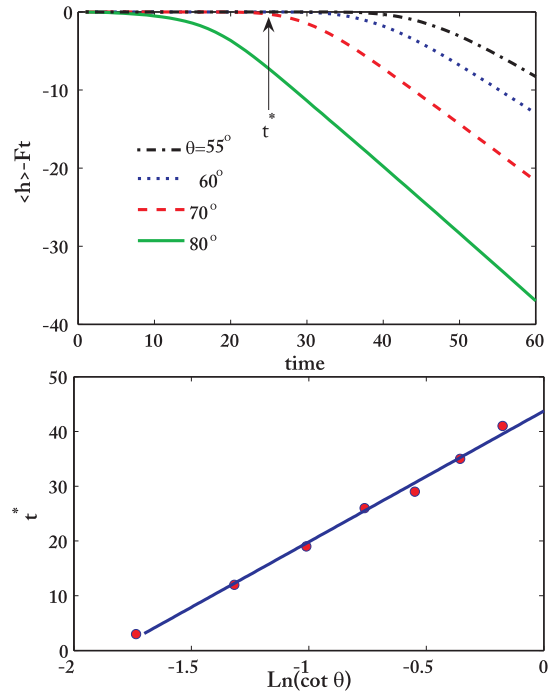}
  \caption{Top plot: evolution of the quantity $v=<h>-Ft$ as a function of time for $\theta=80^o$, $70^o$, $60^o$ and
$55^o$, and for $\nu=K=0.2$. Bottom plot: the crossover time $t^*$
as a function if the incident angle showing the logarithmic
dependence of $t^*$ on $\mu=cot\theta$.}\label{meanHeight}
\end{figure}  \\
We integrated equation (\ref{fullmbeeq}) numerically using
Adams-Bashforth scheme \cite{Hoffman} and imposed periodic
boundary conditions. At each time iteration, the function $S$ is
determined at each point of the profile, before the height is
updated. The initial surface is perturbed with a white noise. The
numerical integration of (\ref{fullmbeeq}) revealed the existence
of an incident angle $\theta^*$ below which the shadowing effect
is irrelevant. This critical angle is mainly determined in the
linear regime and is dependent on the ratio $\nu/K$. The smaller
this ratio, the larger is the critical angle. Typically,
$\theta^*$ is in the range 45$^o$-50$^o$ for $\nu/K$ varying from
10 to 1. To monitor the effect of oblique incidence on growth we
followed the evolution of the mean height of the evolving profile,
a quantity which is accessible experimentally with the help of
scanning probe microscopes. Figure \ref{meanHeight} shows the plot
of $v=<h>$-$Ft$ versus time( here $<h>$ is the mean height), for
$\theta=80^o$, $70^o$, $60^o$ and $55^o$, and for $\nu=K=0.2$ and
$F=1$. For incident angles larger than the critical angle, one can
distinguish two regimes: the first one where $<h>=Ft$
corresponding to the linear regime; the second corresponding to a
growth phase where the shadowing influence becomes relevant. This
clear distinction identifies a time $t^*$ separating the two
regimes. For angles smaller than the critical angle, the time
$t^*$ becomes extremely long and the separation of the linear and
the shadowing regimes is no longer clear due to the dominance of
the ES non-linearities. In this case, the growth of the surface
profile is mainly determined by diffusion and the ES currents
since equation (\ref{fullmbeeq}) is reduced to the well known MBE
equation. The time $t^*$ can be estimated from geometrical
considerations as follows. During the linear regime the typical
height is $\sigma$ which is the surface width defined as
$\sigma^2=\Sigma_{i}\left(h(i)-<h>\right)^2\sim \exp({\nu
t/l^2})$, where $l$ is the typical distance between mounds given
by $l=2\pi\sqrt{(2K/\nu)}$\cite{Krug02}. If we consider two
neighboring mounds of heights $h_{i-1}$ and $h_{i}$ (i is the
position on the profile) then the mound $(i-1)$ shadows the mound
$i$  when $cot\theta<\frac{h_{i-1}-h_i}{l}=\frac{\partial
h}{\partial x}\sim\sigma/l$. This gives:
\begin{equation}\label{crossover}
    t > t^* \sim \frac{l^2}{\nu} \mbox{ln}(cot\theta)
\end{equation}
Figure \ref{meanHeight}-b shows the logarithmic dependence of
$t^*$ on $\mu=cot\theta$.  The time $t^*$ was identified as the
time when the quantity $v=<h>-Ft$ drops from zero to negative
values as clearly shown in figure \ref{meanHeight}.

\begin{figure}
  \includegraphics[width=9 cm]{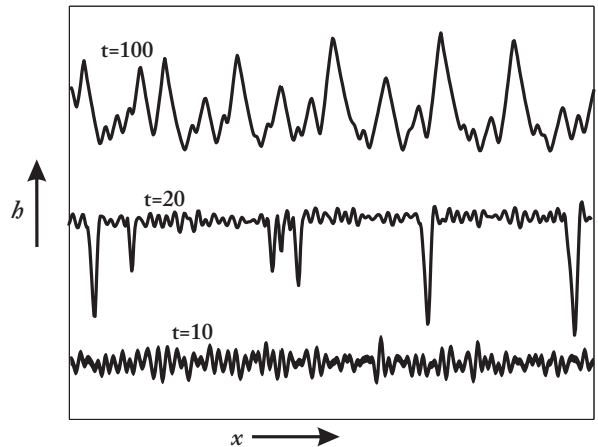}\\
  \caption{Portion of the evolving profile for three different times: t=10, 20 and 100 for an incident angle of
$\theta=80^o$. At t=10, the growth is dominated by the linear
instability. The intermediate phase (t=20) is characterized by the
development of deep grooves  on the surface as the shadowing
starts to become relevant. Later on (t=100), the deep grooves
disappear and the surface settles to a morphology dominated by the
ES effect and shadowing, characterized by developing mounds and
coarsening. The parameters used here are $\nu=0.2$, $K=0.2$ and
$F=1$. The system size was $N=500$.}\label{height}
\end{figure}

\begin{figure}
  \includegraphics[width=9cm]{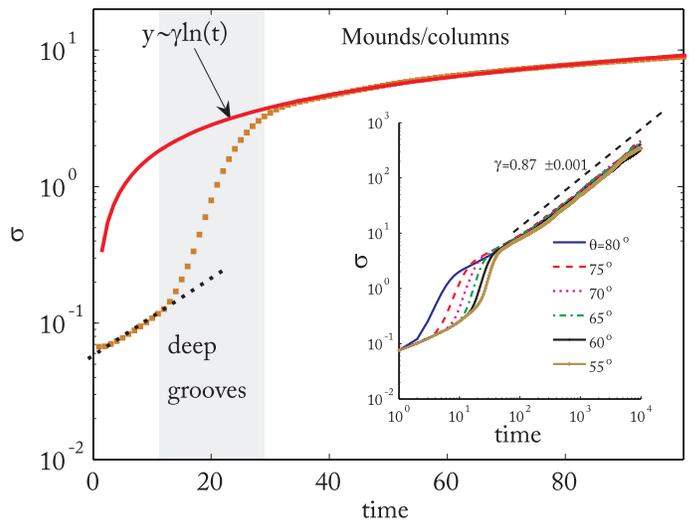}
  \caption{Time evolution of surface width $\sigma$ showing the clear distinction
between three growth phases, for an incident angle of
$\theta=75^o$ and a system size $N=500$. The parameters used here
are $\nu=.1$, $K=.2$ and $F=1$. The early time growth is well
predicted by the linear theory i.e. $\sigma\sim\exp(\nu t/l^2)$.
The deep groove phase (indicated by the shaded area) induces a
sudden increase of the surface width, followed by a growth phase
where it evolves following a power law e.g. $\sigma \sim
t^\gamma$. The inset shows a log-log plot of $\sigma$ vs $t$ for
long integration times. The power law exponent $\gamma$ was
determined by non-linear regression fit giving a value
$\gamma=0.87 \pm 0.001$. The result was averaged over 20
realizations.}\label{rms_h}
\end{figure}

\indent Beyond the time $t^*$, nonlinearities caused by shadowing
and the ES barrier become significant.
 Straight  after the linear regime and before non-linearities
become fully developed, the height $h$ goes through a phase where
deep grooves form. This is because valleys become shadowed and
stop growing whilst the mound tips continue to grow. At this
stage, the ES effect is still weak. As soon as the ES effect
starts to be relevant, the deep grooves acquire a mean velocity
due to the ES non-linearities, and the grooves morphology
disappear, resulting in a morphology dominated by asymmetric
mounds or columns at glancing angles.
 Figure \ref{height} shows an example of the
evolution of the profile's morphology, demonstrating the three
phases for the following parameters : $\theta=80^o$, $\nu=0.1$,
$K=0.2$ and $F=1$. This morphological
behavior is true for any $\theta>\theta^*$.\\
\indent These observations can be quantified by considering the
surface width $\sigma$. In figure \ref{rms_h} the time evolution
of this quantity is showing the clear distinction between the
above mentioned growth phases. The parameters used here are
$\nu=.1$, $K=.2$ and $F=1$. The early time growth is well
predicted by the linear theory i.e. $\sigma=\exp(\nu t/l^2)$. The
deep groove phase (indicated by the shaded area in figure
\ref{rms_h}) induces a sudden increase of the surface width; this
phase is followed by a phase where the surface width evolves
following a power law e.g. $\sigma \sim t^\gamma$.
In the absence of shadowing, the scenario predicted by equation
(\ref{fullmbeeq}) is as follows: after the linear phase where the
surface profile undergoes an exponential growth and where mounds
form, a coarsening phase develops. During this phase, mounds
coarsen and the typical mound size $\lambda$ follows a power law
increase in time i.e. $\lambda\sim t^\beta$, where
$\beta=1/3$\cite{Krug02}. In the presence of shadowing, this
scenario remains similar. Indeed the coarsening is still
persistent but evolves more rapidly than in the absence of
shadowing. To show this, we performed long time integration of
equation (\ref{fullmbeeq}) and computed the height autocorrelation
function $C(r)=<h(x+r)h(x)>_x$. The mounds lateral size $\lambda$
is given by the zero crossing of $C(r)$.
\begin{figure}
  \includegraphics[width=9cm]{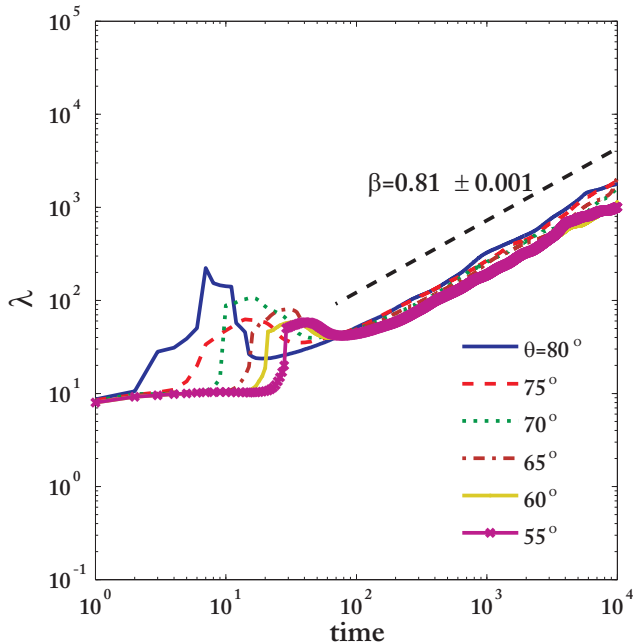}\\
  \caption{Log-log plot of the mound size $\lambda$ versus time for  incident angles
 $\theta=80^o$, $75^o$, $70^o$, $65^o$, $60^o$ and $55^o$. The power law exponent was determined by regression fit which gives the value of
 $\beta=0.81 \pm 0.001$.}\label{corr}
\end{figure}
Figure \ref{corr} displays a log-log plot of the mounds lateral
size versus time for different incident angles. The size of the
mounds follows a power law behavior $\lambda \sim t^\beta$ with
the exponent $\beta=0.81 \pm 0.001$.  Notice that the value of the
exponent is the same for all angles (larger than the critical
angle). This tells us the coarsening is faster than in growth in
normal incidence where the value of the exponent is 1/3, which is
re-confirmed by the integration
of equation (\ref{fullmbeeq}) for angles smaller than $\theta^*$.\\
\indent The draw back of the model (\ref{fullmbeeq}) is it does
not include faceting to account for crystallographic effects and
it is restricted to solid-on-solid (SOS) picture. Shim and Amar
\cite{Amar} used kinetic Monte Carlo method which included
faceting to simulate homoepitaxial growth of metal(100) under
oblique incidence and found that many aspects of the surface
morphology can be explained by purely geometrical effects induced
by shadowing and demonstrated the appearance of ripples and rods
with sides dominated by (111)-oriented facets. Model
(\ref{fullmbeeq}), although it is one dimensional, qualitatively
shares some features observed in these simulations; in particular
the power law behavior of the surface width and the mounds size as
well as the development of asymmetric mounds. In addition, model
(\ref{fullmbeeq}) provided an insight onto the interplay between
different growth mechanisms involved in the growth process.
Another advantage is that it can simulate long time behavior of
the surface morphology without exorbitant computer resources. A
natural progression towards simulations which can directly be
compared to experimental observations is to extend model
(\ref{fullmbeeq}) to two dimensions. This will have an impact on
the experimental design of naturally evolving nanostructures
without resorting to expensive lithographic methods such as
electron beam lithography.

% ----------------------------------------------------------------


\begin{thebibliography}{10}
\bibitem{Villain} A. Pimpinelli and J. Villain 1998 {\it Physics of
 crystal growth} (Cambridge University press)
 \bibitem{Krug02} J. Krug, Physica A \textbf{313}, 47, (2002).
\bibitem{Ehrlish} G. Ehrlich and F.G. Hudda, J. Chem. Phy.
\textbf{44}, 1039 (1966).
\bibitem{Schwoebel} R. L. Schwoebel, J. Appl. Phys. \textbf{40},
614 (1969).
\bibitem{VanDijken} S. van Dijken, L.C. Jorritsma, and B. Poelsema, Phys. Rev. Lett. \textbf{82}, 4038
(1999); S. van Dijken, L. C. Jorritsma  and B. Poelsema Phys. Rev.
B \textbf{61}, 14047 (2000).
\bibitem{Amar} Y. Shim and J. G. Amar, Phys. Rev. Lett,
\textbf{98}, 046103 (2007).

\bibitem{Karunasiri} R. P. Karunasiri,
R. Bruinsma, and J. Rudnick, Phys. Rev. Lett. \textbf{62},
788(1989). R. P. Karunasiri, R. Bruinsma, and J. Rudnick, Phys.
Rev. Lett. \textbf{62}, 2767 (1989).

\bibitem{Bales} G. S. Bales, R. Bruinsma, E. A. Eklund, R. P. U.
Karunasiri, J. Rudnick and A. Zangwill,  Science, \textbf{249},
264 (1990).



\bibitem{Politi2000} P. Politi, G. Grenet, A. Marty, A. Ponchet and J.
Villain, Phys. Rep. \textbf{324}, 271, (2000).

\bibitem{Stroscio} J. A. Stroscio, D. T. Pierce, M. D. Stiles, A.
Zangwill and L. M. Sander, Phys. Rev. Lett. \textbf{75}, 4246,
(1995).
\bibitem{Politi98} P. Politi, Phys. Rev. E \textbf{58}, 281, (1998).
\bibitem{Olgilvy} J. A. Olgilvy, {\it Theory of Wave Scattering from Random Rough
Surfaces} (Bristol: Institute of Physics Publishing 1991).

\bibitem{Bourlier2003} C. Bourlier and G. Berginc, waves in random
media, \textbf{13}, 27, (2003).

\bibitem{Wagner_Smith} R. J. Wagner, J. Acoust. Soc. Am. \textbf{41},
138, (1967). B. G. Smith, IEEE Trans. Antennas Propag.
\textbf{AP-5}, 668, (1967).

\bibitem{Brokelman} R. A. Brokelman and T. Hagfors, IEEE Trans. Antennas
Propag.\textbf{14}, 621, (1966).

\bibitem{Hoffman} Joe D. Hoffman 2001 {\it Numerical Methods for Engineers and
 Scientists} (CRC press)
\end{thebibliography}
\end{document}